\documentstyle[twoside,fleqn,espcrc2,epsfig]{article}

\newcommand{\AmS}{{\protect\the\textfont2
  A\kern-.1667em\lower.5ex\hbox{M}\kern-.125emS}}

\title{The Deep-Inelastic Structure Functions of $\pi$ and $\rho$ 
Mesons\thanks{Poster presented by G. Schierholz}}

\author{
    C.~Best$^{\rm a}$,
    M.~G\"ockeler$^{\rm b}$, 
    R.~Horsley$^{\rm c}$, 
    H.~Perlt$^{\rm d}$,
    P.~Rakow$^{\rm e}$,
    A. Sch\"afer$^{\rm b}$,
    G.~Schierholz$^{\rm e,f}$,
    A.~Schiller$^{\rm d}$
    and S.~Schramm$^{\rm g}$ \\[1em]
    $^{\rm a}$ Institut f\"ur Theoretische Physik, J.~W.~Goethe
    Universit\"at, D-60054 Frankfurt \\[0.5em]
    $^{\rm b}$ Institut f\"ur Theoretische Physik, 
    Universit\"at Regensburg, D-93040 Regensburg \\[0.5em]
    $^{\rm c}$ Institut f\"ur Physik, Humboldt-Universit\"at zu Berlin, 
    D-10115 Berlin\\[0.5em]
    $^{\rm d}$ Institut f\"ur Theoretische Physik, Universit\"at Leipzig, 
    D-04109 Leipzig\\[0.5em]
    $^{\rm e}$ Deutsches Elektronen-Synchrotron DESY, Institut f\"ur
    Hochenergiephysik und HLRZ,\\
    \hspace*{0.174cm} D-15735 Zeuthen\\[0.5em]
    $^{\rm f}$ Deutsches Elektronen-Synchrotron DESY, 
    D-22603 Hamburg\\[0.5em]
    $^{\rm g}$ Gesellschaft f\"ur Schwerionenforschung GSI,
    D-64220 Darmstadt\\[0.5em]}
       
\begin{document}

\begin{abstract}
We compute the lower moments of the structure functions of $\pi$ and $\rho$.
Of particular interest are the spin-dependent structure functions of the 
$\rho$ as they give new information about quark binding effects.
\end{abstract}

\maketitle

\setcounter{footnote}{0}

\section{INTRODUCTION}
\label{intro}

Much of our knowledge about QCD and the structure of hadrons has been 
derived from deep-inelastic scattering experiments. While most work to date
has been on the nucleon, there are proposals to investigate the structure
functions of mesons~\cite{prop}. In this talk we shall consider $\pi$ and
$\rho$ targets.

No polarization is possible for the $\pi$, so we have two structure functions:
$F_1$ and $F_2$. The $\rho$, being a spin 1 particle, has eight structure
functions~\cite{jaffe}: $F_1$, $F_2$, $g_1$, $g_2$, $b_1$, $b_2$, $b_3$ and
$b_4$. The structure functions $b_3$ and $b_4$ receive contributions
from twist-four (and higher) operators only and so will not concern us here.
We shall also not consider the structure function $g_2$,
which is least likely that it will be measured.

Writing
\begin{eqnarray*}
M_n(f) = \int_0^1 \mbox{d}x x^{n-1} f(x), & &
\end{eqnarray*}
we find for the moments of the $\pi$ structure functions 
\begin{eqnarray*}
2M_n(F_1) &=& c_n^{(1)} v_n, \\
M_{n-1}(F_2) &=& c_n^{(2)} v_n,
\end{eqnarray*}
while the moments of the $\rho$ structure functions are given by
\begin{eqnarray*}
2M_n(F_1) &=& c_n^{(1)} a_n,  \\
M_{n-1}(F_2) &=& c_n^{(2)} a_n, \\
2M_n(b_1) &=& c_n^{(1)} d_n,  \\
M_{n-1}(b_2) &=& c_n^{(2)} d_n, \\
2M_n(g_1) &=& c_n^{(3)} r_n,
\end{eqnarray*}
where the $c_n$'s are the Wilson coefficients, and $v_n$, $a_n$, $d_n$ and 
$r_n$ are the operator matrix elements. The latter derive from two 
operators,
\begin{eqnarray*}
{\cal O}_{\mu_1\cdots\mu_n} = \frac{1}{2^{n-1}} \bar{\psi}\gamma_{\mu_1}
\mbox{i}\stackrel{\leftrightarrow}{D}_{\mu_2} \cdots 
\mbox{i}\stackrel{\leftrightarrow}{D}_{\mu_n}\psi - \mbox{Tr}, & & 
\end{eqnarray*}
\vspace*{-0.5cm}
\begin{eqnarray*}
{\cal O}^5_{\mu_1\cdots\mu_n} = \frac{1}{2^{n-1}} \bar{\psi}\gamma_{\mu_1}
\gamma_5 \mbox{i}\stackrel{\leftrightarrow}{D}_{\mu_2} \cdots 
\mbox{i}\stackrel{\leftrightarrow}{D}_{\mu_n}\psi - \mbox{Tr}, & &
\end{eqnarray*}
and are given by
\begin{eqnarray*}
\langle\pi|{\cal O}_{\{\mu_1\cdots\mu_n\}}|\pi\rangle = 2 v_n 
p_{\mu_1} \cdots p_{\mu_n} - \mbox{Tr} & &
\end{eqnarray*}
and
\begin{eqnarray*}
\lefteqn{\langle\rho,\lambda|{\cal O}_{\{\mu_1\cdots\mu_n\}}
|\rho,\lambda'\rangle 
= 2 S [a_n p_{\mu_1} \cdots p_{\mu_n}\delta_{\lambda \lambda'}}\\
& &+d_n (\epsilon_{\mu_1}^{* \lambda}\epsilon_{\mu_2}^{\lambda'} 
- \frac{1}{3} p_{\mu_1}p_{\mu_2} \delta_{\lambda \lambda'} )
p_{\mu_3}\cdots p_{\mu_n}], 
\end{eqnarray*}
\vspace*{-0.5cm}
\begin{eqnarray*}
\lefteqn{\langle\rho,\lambda|{\cal O}^5_{\{\mu_1\cdots\mu_n\}}
|\rho,\lambda'\rangle = \frac{2\mbox{i}}{m_\rho^2} 
S [r_n \epsilon_{\rho\sigma\tau\mu_1}
\epsilon_{\rho}^{* \lambda}\epsilon_{\sigma}^{\lambda'}}\\
& & \times p_\tau p_{\mu_2}\cdots
p_{\mu_n}],
\end{eqnarray*}
where $S$ symmetrizes and subtracts traces.

In parton model language
\begin{eqnarray*}
b_1(x)=\frac{1}{2} [ q^0(x) - q^1(x)], & &
\end{eqnarray*}
where $q^m(x)$ is the probability to find a quark of fractional momentum $x$
in a $\rho$ with spin projection $m$ relative to its momentum. If the quarks 
are in a pure $s$-wave state, we would expect $b_1$ to vanish. The other
structure functions have more or less the same interpretation as in case of
the nucleon.

\section{LATTICE CALCULATION}

We have generated 500 quenched gauge field configurations on a
$16^3 32$ lattice at $\beta = 6.0$. On these configurations we have
computed the matrix elements $v_n$, $a_n$, $d_n$ and $r_n$, using Wilson 
fermions. The calculations are done for three hopping parameters, $\kappa =
0.1515$, 0.1530 and 0.1550, corresponding roughly to quark masses of
190, 130 and 70 MeV, respectively.  

\begin{figure}[bht]
\begin{centering}
\epsfig{figure=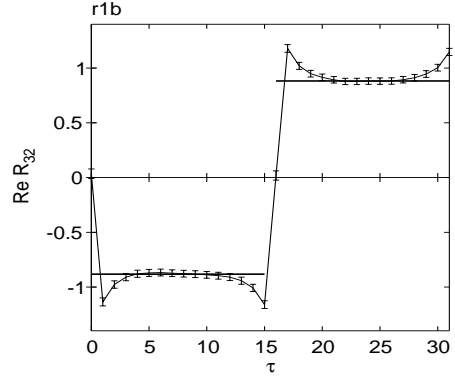,height=6.0cm,width=7.0cm}
\vspace{-0.5cm}
\caption{The ratio $\mbox{Re}R_{32}$ for the polarized operator $r_{1b}$. 
The subscript $\scriptsize b$ stands for a particular 
representation~\protect\cite{best}. The
sink is placed at $t=16$, so that we can use both plateaus.}
\vspace{-0.50cm}
\end{centering}
\end{figure}

The matrix elements can be obtained from the ratio of three- to two-point
functions~\cite{prd}
\begin{eqnarray*}
\frac{\langle\eta(t) {\cal O}(\tau) \eta^\dagger(0)\rangle}
{\langle\eta(t) \eta^\dagger(0)\rangle} = R & &
\end{eqnarray*}
for $t \gg \tau \gg 0$, where $\eta = \pi, \rho$ is the source/sink. For the
$\pi$ we have
\begin{eqnarray*}
R \propto \langle\pi|{\cal O}|\pi\rangle, & &
\end{eqnarray*}
while for the $\rho$ we have
\begin{eqnarray*}
R_{ij} \propto \sum_{\lambda,\lambda'} 
\epsilon^\lambda_i\epsilon^{*\lambda'}_j 
\langle\rho,\lambda|{\cal O}|\rho,\lambda'\rangle. & &
\end{eqnarray*}
A typical such ratio is shown in Fig.~1.

The operators will give in general divergent results as the lattice 
spacing goes to zero. They must be renormalized in the same way the Wilson
coefficients must be renormalized. We define finite operators renormalized
at the scale $\mu$ by ($a$ being the lattice constant)
\begin{eqnarray*}
{\cal O}(\mu) = Z_{\cal O}((a\mu)^2, g(a)) {\cal O}(a).
\end{eqnarray*}
The renormalization constants $Z$ have been computed to one loop order in
perturbation theory~\cite{npb}.

\section{RESULTS}

We work at a scale of $\mu^2 = (1/a)^2 \approx 4 \mbox{GeV}^2$. We shall 
present our results as a series of plots. In Fig.~2 we show the moments
of the $\pi$ structure function $F_1$.
For the lowest moment, $<x>$, we find that it is somewhat larger than the 
phenomenological result. The higher moments are in better agreement. This is
similar to what has been found for the nucleon structure functions~\cite{prd}. 

The unpolarized $\rho$ structure function looks very similar to the $\pi$
structure function, so that we will not discuss it here.
In Fig.~3 we show the moments of the polarized $\rho$ structure function
$g_1$. The lowest moment $r_1$ indicates that the valence quarks carry
about $60\%$ of the total spin of the $\rho$. A similar calculation
for the nucleon gave a quark spin fraction of the same value~\cite{prd}.

\clearpage
\begin{figure}[t]
\vspace{-0.8cm}
\begin{centering}
\epsfig{figure=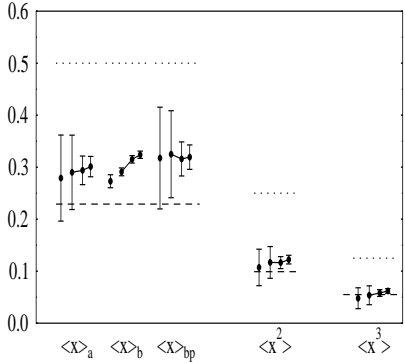,height=6.0cm,width=7.0cm}
\vspace{-0.7cm}
\caption{The moments of the $\pi$ structure function. The subscripts 
$\scriptsize a, b, bp$ stand for different 
representations~\protect\cite{best}. For 
each moment the heaviest quark mass is on the right. The leftmost value 
is the extrapolation to the chiral limit. The heavy quark mass limit is 
given by the dotted lines, while the experimental value~\protect\cite{sutton} 
is given by the dashed lines.}
\end{centering}
\vspace{-0.50cm}
\end{figure}

In Fig.~4 we show the moments of the polarized $\rho$ structure function 
$b_1$. The lowest moment turns out to be positive and surprisingly large,
albeit with large statistical errors. Perhaps this indicates that the
valence quarks have a substantial orbital angular momentum. 

\begin{figure}[h]
\vspace{-1.0cm}
\begin{centering}
\epsfig{figure=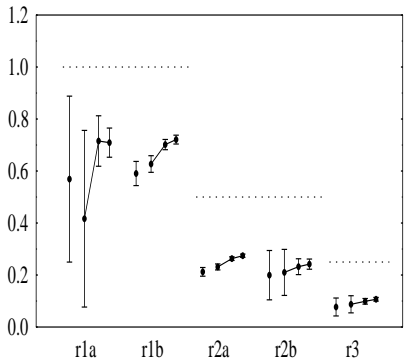,height=6.0cm,width=7.0cm}
\vspace{-0.7cm}
\caption{The moments $r_n$ of the polarized $\rho$ structure function $g_1$.
The subscripts $\scriptsize a, b$ stand for different representations.
The lattice data are plotted in the same way as in Fig.~2.}
\end{centering}
\vspace{-3.0cm}
\end{figure}

\newpage
\begin{figure}[tbh]
\vspace{-0.75cm}
\epsfig{figure=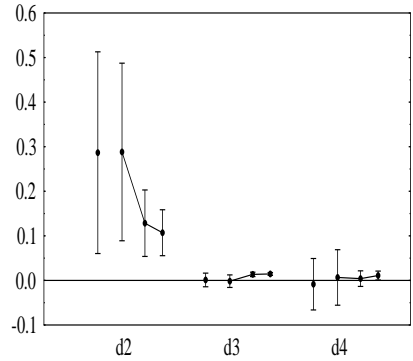,height=6.0cm,width=7.0cm}
\vspace{-0.7cm}
\caption{The moments $d_n$ of the polarized $\rho$ structure function $b_1$.
The lattice data are plotted in the same way as in Fig.~2.}
\vspace{-0.50cm}
\end{figure}

\section*{ACKNOWLEDGMENTS}
\label{acknowledgements}

The numerical calculations were performed on the Quadrics parallel
computers at Bielefeld University. Financial support by the Deutsche Forschungsgemeinschaft is gratefully acknowledged.

\end{document}